\documentclass[twocolumn,secnumarabic,amssymb,nobibnotes,aps,prl]{revtex4-1}

\usepackage[dvips]{graphicx}
\usepackage{bm}
\usepackage{latexsym}
\usepackage{amsmath}
\usepackage{amssymb}
\usepackage{color}

\makeatletter
\renewcommand*{\p@subsection}{}
\renewcommand*{\p@subsubsection}{}
\makeatother
\begin{document}
\title{Quantum Oscillations of The Positive Longitudinal
Magnetoconductivity:\\  a Fingerprint for Identifying  Weyl Semimetals}
\author{Ming-Xun Deng$^{1,2}$}
\author{G. Y. Qi $^{1}$}
\author{R. Ma $^{3}$}
\author{R. Shen$^{1,4}$}
\author{L. Sheng$^{1,4}$}
\email{shengli@nju.edu.cn}
\author{D. Y. Xing$^{1,4}$}
\email{dyx@nju.edu.cn}
\affiliation{$^{1}$ National Laboratory of Solid State Microstructures and Department of
Physics, Nanjing University, Nanjing 210093, China}
\affiliation{$^{2}$Laboratory of Quantum Engineering and Quantum Materials, ICMP and SPTE,
South China Normal University, Guangzhou 510006, China}
\affiliation{$^{3}$ Jiangsu Key Laboratory for Optoelectronic Detection of Atmosphere and
Ocean, Nanjing University of Information Science and Technology, Nanjing
210044, China }
\affiliation{$^{4}$ Collaborative Innovation Center of Advanced Microstructures, Nanjing
University, Nanjing 210093, China}
\date{\today }

\begin{abstract}
Weyl semimetals (WSMs) host charged Weyl fermions
as emergent quasiparticles. We develop a unified analytical theory for the anomalous
positive longitudinal magnetoconductance (LMC) in a WSM,
which bridges the gap  between the classical
and ultra-quantum approaches. More interestingly,
the LMC is found to exhibit periodic-in-$1/B$ quantum oscillations, originating
 from the oscillations of the nonequilibrium
 chiral chemical potential. The quantum oscillations,
 superposed on the positive LMC, are a remarkable fingerprint of a WSM phase with chiral anomaly,
 whose observation is a valid criteria for identifying a WSM material. In fact, such quantum oscillations
 were already observed by several experiments.

\end{abstract}

\maketitle

Weyl semimetals (WSMs), whose low-energy excitations are Weyl fermions~\cite{Phys.56.330}
carrying charges, have recently spurred intensive and innovative
research in the field of condensed matter
physics~\cite{PhysRevB.83.205101,PhysRevLett.107.186806,PhysRevLett.107.127205,PhysRevB.84.075129,PhysRevB.85.035103,PhysRevLett.111.246603,HOSUR2013857}.
The ultrahigh mobility and spectacular transport properties
of the charged Weyl fermions can find applications in high-speed electronic circuits and  computors~\cite{Ali2014,Shekhar2015,PhysRevX.4.031035}.
The low-energy spectrum of a WSM forms non-degenerate three-dimensional (3D) Dirac cones
around isolated degenerate band touching points, referred to as Weyl points~\cite{PhysRevB.83.205101}. Weyl points
with opposite chiralities, playing
the parts of the source and sink of Berry curvature in  momentum space,
always come in pairs~\cite{Volovik2003,Nielsen389}. The appearance of
the Weyl points requires breaking either the spatial inversion or time-reversal symmetry.
Weyl points with opposite chiralities in momentum
space are connected by the
nonclosed Fermi arc surface states~\cite{PhysRevB.83.205101}.


The WSM state was first realized experimentally in TaAs~\cite{Xu613,PhysRevX.5.031013,Nat.Phys.11.724}, following
the theoretical predictions~\cite{Nat.Commun.6.7373,PhysRevX.5.011029}, and
later in several different compounds~\cite{Nat.Phys.11.728,Shekhar2015,Xiong413,ncomms10137,ncomms13142,LvPRL096603,PhysRevX.5.031023,ncomms10301,ncomms10735,PRB041104R,FP127203,China657406}.
WSMs display many anomalous transport properties, such as
positive longitudinal magnetoconductance (LMC), optical gyrotropy~\cite{PRB161110},
planar Hall effect~\cite{PRL176804}, all of which are induced by the chiral anomaly~\cite{Nielsen389},
and nonlocal quantum oscillations
of the Fermi arc surface
states~\cite{Nat.Commun.5.5161}. The chiral anomaly, also termed
as the Adler-Bell-Jackiw anomaly, means the violation
of the separate number conservation laws of Weyl fermions of different chiralities.
Parallel electric and magnetic fields can pump Weyl fermions between
Weyl valleys of opposite chiralities, and create a population
imbalance between them,
therefore resulting in a positive LMC (or negative magnetoresistance).
 The anomalous LMC, as an exotic macroscopic quantum phenomenon,
has been attracting intense experimental~\cite{Shekhar2015,Xiong413,ncomms10137,ncomms13142,LvPRL096603,PhysRevX.5.031023,ncomms10301,ncomms10735,PRB041104R,FP127203,China657406} and theoretical~\cite{PhysRevLett.111.246603,PhysRevB.85.241101,PhysRevB.88.104412,PhysRevB.91.245157,PRB165101} interest.

In order to identify a WSM material, the ARPES experiments were
used to directly observe the Weyl nodes and Fermi arcs~\cite{Xu613,PhysRevX.5.031013,Nat.Phys.11.724,Nat.Phys.11.728}.
However, the ARPES identification is sometime limited by spectroscopic resolutions.
Another widely-employed method
is to measure the positive LMC induced by the chiral anomaly~\cite{Shekhar2015,Xiong413,ncomms10137,ncomms10735,ncomms13142,LvPRL096603,
PhysRevX.5.031023,PRB041104R,ncomms10301,FP127203,China657406}.
The observation of the positive LMC is only a necessary condition for identifying the WSM phase, but not a sufficient condition.
On the other hand, in the classical limit, $\vert E_{F}\vert\gg\hbar\omega_{c}$, an analytical formula for the
anomalous LMC was derived in Refs.\ ~\cite{PhysRevB.88.104412,PhysRevB.91.245157}, yielding $\Delta\sigma(B)\propto (B/E_{F})^2$
with $E_F$ the Fermi energy, $\omega_{c}$ the cyclotron frequency, and $B$ the strength of magnetic field. In the opposite ultra-quantum
limit, $\vert E_{F}\vert\ll\hbar\omega_{c}$, it was shown
independently~\cite{PhysRevB.88.104412,PRB165101} that $\Delta\sigma(B)\propto B$.
While the above quadratic or linear field dependence of the positive LMC was suggested as
an additional signature of a WSM, in experiments the simultaneous presence of
negative field-dependent magnetoconductivity due to weak anti-localization~\cite{Shekhar2015,Xiong413,ncomms10137,ncomms10735,ncomms13142,LvPRL096603,
PhysRevX.5.031023,PRB041104R,ncomms10301,FP127203,China657406}
 often makes comparison of experimental data with the theories equivocal.
Theoretical investigation of the anomalous LMC in the intermediate regime between the classical
and ultra-quantum limits, where $\vert E_{F}\vert$ and $\hbar\omega_{c}$
 are comparable to each other, is still absent. It is urgent to develop a unified theory across the two limits by taking into account the interplay between the chiral anomaly and Landau
quantization, and in particular to seek out a fingerprint identification
of a WSM material based on transport measurements.

The main purpose of this work is twofold. First, integrating the Landau  quantization
with Boltzmann equation, we derive a unified analytical formula
for the anomalous LMC in a WSM, which is applicable to a broad range
from the classical to ultra-quantum limit. It
recovers the known results in the two opposite limits. More interestingly,
we find that the anomalous positive LMC displays periodic-in-1/B quantum oscillations,
originating from the oscillations of the nonequilibrium chiral chemical potential.
Second, we propose that the quantum oscillations superposed on the positive LMC are
an important fingerprint for identifying a WSM material with chiral anomaly,
which was not disclosed in previous theories.
Unlike the quadratic or linear field dependence, the quantum oscillations
of the anomalous LMC will not be concealed by the
presence of negative
magnetoconductivity due to weak anti-localization.
In fact, such quantum oscillations were already observed by
several experimental works, e.g., see Figs.\ 3(a,b) in Ref.~
\cite{ncomms10137}, Fig.\ 2(d) in Ref.~\cite{ncomms10735},
and Fig.\ 3(d) in Ref.~\cite{China657406}.


\begin{figure}[ptb]
\centering
\includegraphics[scale=0.80]{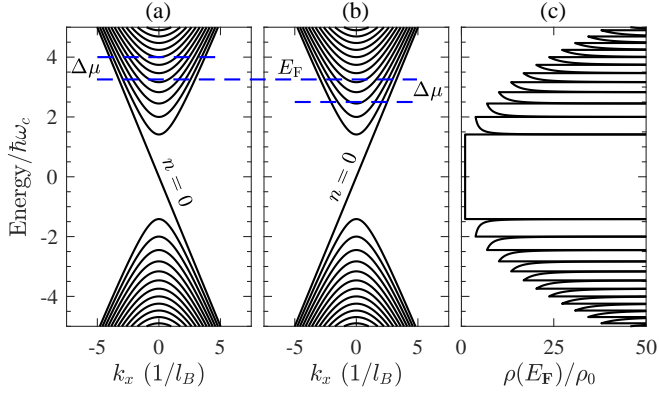}
\caption{ The LLs in (a) $\chi=+$, and (b) $\chi=-$ Weyl valleys. (c) The DOS of the Weyl fermions (horizontal axis) as a function of normalized Fermi energy $E_{F}/\hbar\omega_{c}$
(vertical axis).
Blue dashed lines in (a,b) are an
enlarged illustration of the effect of chiral anomaly. In the steady state,
the local chemical potentials in the two valleys shift upward and downward relative to $E_{F}$, respectively,
by an equal amount $\Delta\mu$, leading to a transfer of charged fermions between the two valleys. }%
\label{FigLLs}%
\end{figure}

Let us start by considering a 3D WSM, which has two Weyl points
with opposite chiralities,
labelled by $\chi=\pm$. When a magnetic
field $\mathbf{B}=(B,0,0)$ is applied along the $x$ direction,
the continuum Hamiltonian for low-energy  electrons
in a Weyl valley reads
\begin{equation}
H_{\chi}({\mathbf{k}})=\chi\upsilon_{\mathrm{F}}\left(\hbar\mathbf{k}
+e\mathbf{A}\right)\cdot
\mbox{\boldmath{$\sigma$}}\ ,\label{H0}%
\end{equation}
where the electron charge is taken to be $-e$, $\upsilon_{\mathrm{F}}$
is the Fermi velocity, $\mbox{\boldmath{$\sigma$}}=(\sigma_{x},\sigma_{y},\sigma_{z})$
are the Pauli matrices, $\mathbf{k}$ is the wave vector, and $\mathbf{A}$ is the vector
potential defined by $\mathbf{B}=\nabla\times \mathbf{A}$.
Its energy spectrum  can be solved exactly, yielding
\begin{equation}
\varepsilon_{n}^{\chi}(k_{x})=%
\begin{cases}
-\chi\hbar\upsilon_{\mathrm{F}}k_{x} & n=0\\
\mbox{sgn}(n)\sqrt{2|n|(\hbar\omega_{c})^{2}+(\hbar\upsilon_{\mathrm{F}}k_{x})^{2}} &
n\neq 0
\end{cases}
\label{disper}%
\end{equation}
with $\ell_{B}=\sqrt{\hbar /eB}$ as the magnetic length and $\omega_{c}%
=\upsilon_{\mathrm{F}}/\ell_{B}$.
The degeneracy of each Landau level (LL) is equal to $\Omega_{n}^{\chi}=1/2\pi l^{2}_{B}$
per unit cross-section.
The longitudinal group velocity for the $n$-th LL is given by
\begin{equation}
\upsilon_{x,n}^{\chi}(k_{x})=\frac{\partial \varepsilon_{n}^{\chi}(k_{x})}
{\hbar\partial k_{x}}=\begin{cases}
-\chi\upsilon_{\mathrm{F}} & n=0\\
{\hbar\upsilon_{\mathrm{F}}^{2}k_{x}}/{\varepsilon_{n}^{\chi}(k_{x})} &
n\neq 0
\end{cases}
\ .\label{vx}%
\end{equation}
The LLs are plotted in Figs.\ \ref{FigLLs}(a) and \ref{FigLLs}(b), whose slopes correspond to group velocities $\upsilon_{x,n}^{\chi}(k_{x})$.
In each Weyl valley, the $n=0$ LL
is chiral, manifesting the chirality of the Weyl point, and all $n\neq 0$ LLs are achiral.

The two valleys have the identical density of states (DOS).
The DOS at $E_{F}$ of a single valley is given by
\begin{eqnarray}
\rho(E_{F})&=&\sum\limits_{n}\frac{1}{2\pi \ell_{B}^2}\int \frac{dk_{x}}{2\pi}
\delta\bigl(E_{F}-\varepsilon_{n}^{\chi}(k_{x})\bigr)\nonumber\\
&=&\rho_{0}\varTheta\ ,\label{DOSs}
\end{eqnarray}
where $\rho_{0}={1}/{2\pi\ell_{B}^2 h \upsilon_{F}}$, and
\begin{equation}
\varTheta=1+2
{\displaystyle\sum\limits_{n=1}^{n_{c}}}
\frac{1}{\lambda_{n}}\ , \label{Thet}
\end{equation}
with $\lambda_{n}=
\sqrt{1-2|n|(\hbar\omega_{c}/E_{\mathrm{F}})^{2}}$. Here, $n_{c}=\mbox{sgn}(E_{F})\mbox{int}\left[E^{2}_{F}/2(\hbar\omega_{c})^2\right]$
is the index of the highest (lowest) LL crossed by the Fermi level
for $E_{F}>0$ ($E_{F}<0$). The DOS $\rho(E_{F})$ is plotted
in Fig.\ \ref{FigLLs}(c). It oscillates strongly with changing
$E_{F}$ due to the oscillating factor $\varTheta$. Whenever the top or bottom
of a LL passes through $E_{F}$, i.e., at
$E_{F}=\mbox{sgn}(n)\sqrt{2\vert n\vert}\hbar\omega_{c}$ with $n
\neq 0$, $\rho(E_{F})$ diverges periodically, exhibiting van Hove singularities.

\begin{figure}[ptb]
\centering
\includegraphics[scale=0.785]{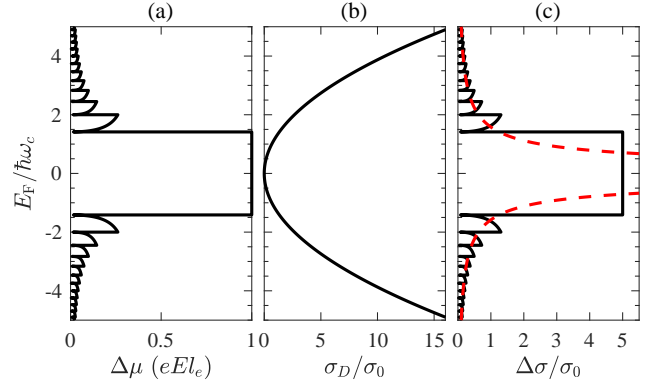}\caption{(a) $\Delta\overline{g}$, (b) $\sigma_D$, and (c) $\Delta\sigma(B)$ as functions of $E_{F}/\hbar\omega_{c}$ for $\tau_{inter} = 5\tau_{intra}$. The red dashed curve plotted in (c) is calculated from the classical formula (\ref{sigA}).
We set the magnetic field to be $B=1T$, and
define $\sigma_{0}=\frac{2e^{2}}{h}\frac{e[B=1T]\upsilon_{F}\tau_{intra}
}{h}$ as the unit of conductivity, for convenience.}%
\label{Figmagto}%
\end{figure}

Upon application of an electric field $\mathbf{E}=(E,0,0)$ along the $x$ direction,
namely, $\mathbf{E}\ {\parallel}\ \mathbf{B}$,
the linear-response steady-state electron distribution
function for
the $n$-th LL in the $\chi$ valley in general takes the form
\begin{equation}
f_{n}^{\chi}(k_{x})=f_{0}(\varepsilon_{n}^{\chi})- f'_{0}(\varepsilon_{n}^{\chi})g_{n}^{\chi}(k_{x})\ ,
\label{def_g}
\end{equation}
where $f_{0}(\varepsilon_{n}^{\chi})=1/[e^{(\varepsilon_{n}^{\chi}-E_{F})/kT}+1]$ is the equilibrium distribution function, $f'_{0}(\varepsilon_{n}^{\chi})=\partial f_{0}(\varepsilon_{n}^{\chi})
/\partial \varepsilon_{n}^{\chi}$, and $g_{n}^{\chi}(k_{x})$ describes the deviation
of the electron distribution function from $f_{0}(\varepsilon_{n}^{\chi})$. The linearized Boltzmann equation reads
\begin{equation}
e\upsilon_{x,n}^{\chi}E
=-\frac{g_{n}^{\chi}(k_{x})-\overline{g}_{\chi}}{\tau_{intra}}%
-\frac{\overline{g}_{\chi}}{\tau_{inter}}\ ,\label{gchi0}%
\end{equation}
with $\overline{g}_{\chi}=\langle
g_{n}^{\chi}(k_{x})\rangle_{\chi}$, where $\tau_{intra}$ and $\tau_{inter}$
stand for the transport relaxation times due to the electron intravalley and intervalley
scattering by impurities. It is assumed that $\tau_{inter}\gg \tau_{intra}$, as the separation of the
Weyl points in the Brillouin zone
usually makes the intervalley scattering much weaker than intravalley
scattering. Here, the average
 $\langle\cdots\rangle_{\chi}$ runs over all electron states at the
 Fermi level in the $\chi$ valley, defined as
\begin{equation}
\langle\cdots\rangle_{\chi}=\frac{\sum_{n}\frac{1}{2\pi\ell_{B}^2}\int \frac{dk_{x}}{2\pi}[-f'_{0}(\varepsilon_{n}^{\chi})](\cdots)}{
\sum_{n}\frac{1}{2\pi\ell_{B}^2}
\int \frac{dk_{x}}{2\pi}[-f'_{0}(\varepsilon_{n}^{\chi})]}\ .\label{def_aver}
\end{equation}

From Eq.\ (\ref{gchi0}), it is easy to obtain a formal solution for $g_{n}^{\chi
}(k_{x})$,
\begin{align}
g_{n}^{\chi}(k_{x}) &  =-e\upsilon_{x,n}^{\chi}E\tau_{intra}+\left(1-\frac{\tau_{intra}}{\tau_{inter}}\right)\overline{g}_{\chi}
\ .\label{fgchi}%
\end{align}
The unknown $\overline{g}_{\chi}$ on the right hand side of Eq.\ (\ref{fgchi})
can be solved self-consistently by averaging the both sides of Eq.\ (\ref{fgchi})
at the Fermi level, yielding
$\overline{g}_{\chi}=-\langle\upsilon_{x,n}^{\chi}\rangle_{\chi}eE\tau_{inter}$.
Here, we go into some details on calculation to understand the role of the chiral anomaly.
It is easy to derive $\langle\upsilon_{x,n}^{\chi}\rangle_{\chi}=\Delta N_{ch}^{\chi}/h\rho(E_{F})$,
which is in inverse proportion to the DOS. Here,
$\Delta N_{ch}^{\chi}
=\frac{1}{2\pi\ell_{B}^{2}}\sum_{\{k_{x}=k_{x,n}^{\chi}(i)\}}\mbox{sgn}(\upsilon_{x,n}^{\chi})$,
 where the summation is to add up the signs
 of the group velocity $\upsilon_{x,n}^{\chi}$ at all the
 intersection points, denoted by  $\{k_{x}=k_{x,n}^{\chi}(i)\}$,
 between the Fermi level $E_{F}$ and LLs in the $\chi$
 valley.
  By definition,
  $\Delta N_{ch}^{\chi}$ is
essentially the number difference between
the right-moving ($\upsilon_{x,n}^{\chi}>0$)
and left-moving ($\upsilon_{x,n}^{\chi}<0$)
channels at $E_F$ in the $\chi$ valley per unit cross-section.
For the achiral $n\neq 0$
LLs, the right-moving and left-moving channels are always in pairs,
making zero contribution to $\Delta N_{ch}^{\chi}$. Therefore,
$\Delta N_{ch}^{\chi}$ equals to the number of chiral channels of the
$n=0$ LL, with its sign determined by the propagating
direction of the chiral channels, $\Delta N_{ch}^{\chi}=
-\chi /2\pi\ell_{B}^{2}$.
Subsequently, we obtain
$\overline{g}_{\chi}=\chi\Delta\mu$, where
\begin{equation}
\Delta\mu=eEl_{e}\frac{1}{\varTheta}\ ,\label{del_g}
\end{equation}
with $l_{e}=\upsilon_{F}\tau_{inter}$. Then Eq.\ (\ref{fgchi}) becomes
 \begin{align}
g_{n}^{\chi}(k_{x}) &  =-e\upsilon_{x,n}^{\chi}E\tau_{intra}+\chi\Delta\mu
\ .\label{fgchi1}%
\end{align}
The term linear in $\tau_{intra}/\tau_{inter}\ll 1$ in Eq.\ (\ref{fgchi}) can now
be omitted.

According to Eq.\ (\ref{def_g}), the second term in Eq.\ (\ref{fgchi1})
corresponds to the nonequilibrium local chemical potentials in
the $\chi=\pm$ valleys relative to $E_{F}$, which are equal
in magnitude but opposite in sign.
Half their difference, $\Delta\mu$,
is called the chiral chemical potential~\cite{PRL031601}.
A nonzero $\Delta\mu$, as illustrated by the blue lines in Fig.\ \ref{FigLLs}(a,b),
indicates that an imbalance
of carrier density is established between the two Weyl valleys.

The electrical current density is given by
\begin{equation}
j_{x}=\frac{-e}{2\pi\ell_{B}^2}%
{\displaystyle\sum\limits_{\chi,n}}
\int\mathbf{\upsilon}_{x,n}^{\chi}(k_{x})g_{n}^{\chi}(k_{x})
\left[-f'_{0}(\varepsilon_{n}^{\chi})\right]\frac{dk_{x}}{2\pi}\ .\label{cds}%
\end{equation}
Substituting Eq.\ (\ref{fgchi1})
into Eq.\ (\ref{cds}), we divide the conductivity  into two parts $\sigma(B)\equiv j_{x}/E=\sigma_{D}+\Delta\sigma(B)$.  The zero-field Drude conductivity is given by
\begin{equation}
\sigma_{D}   =\frac{n_{e}e^2}{\hbar k_{F}}\upsilon_{F}\tau_{intra}\ ,
\end{equation}
with $k_{F}=\vert E_{F}\vert/\hbar \upsilon_{F}$ and $n_{e}=\frac{1}{3\pi^2}k_{F}^{3}$ as the
carrier density. To the leading order in $\tau_{intra}/\tau_{inter}\ll 1$,  $\Delta\sigma(B)\equiv[\sigma(B)-\sigma_{D}]$ can be written as
\begin{equation}
\Delta\sigma(B)=\frac{e}{h E}\left(\Delta N^{-}_{ch}-
\Delta N^{+}_{ch}\right)\Delta\mu\ .\label{Illumin_Sig}
\end{equation}
This formula illuminates the fact that the anomalous LMC comes from
the chiral chemical potential, $\Delta\mu$, driving extra
electrical current to flow through the totally $\left(\Delta N^{-}_{ch}-
\Delta N^{+}_{ch}\right)$ chiral channels in the $n=0$ LLs.
Equation  (\ref{Illumin_Sig}) can be further derived into
\begin{align}
&\Delta\sigma(B)    =\frac{2e^{2}}{h}\frac{eB\upsilon_{F}\tau_{inter}
}{h}\frac{1}{\varTheta}
\ .\label{JxA}
\end{align}
This quantum formula is the central result of our work. $\Delta\sigma(B)$ contains
an oscillating factor $1/\varTheta$, which is introduced in Eq.\ (\ref{Thet}).
Another feature of $\Delta\sigma(B)$
is that it is proportional to
the intervalley relaxation time $\tau_{inter}$, because
the nonequilibrium chiral chemical potential can only relax via
intervalley scattering~\cite{PhysRevB.88.104412,PhysRevB.91.245157}.

In Figs.\ \ref{Figmagto}(a-c), we plot the calculated $\Delta\mu$, $\sigma_{D}$
and $\Delta\sigma(B)$ as functions of the normalized Fermi energy $E_{F}/\hbar\omega_{c}$.
As mentioned above, $\Delta\mu$ is inversely proportional to
the electron DOS. At $E_{F}=\mbox{sgn}(n)\sqrt{2\vert n\vert}\hbar\omega_{c}$ with $n
\neq 0$, where the DOS diverges, $\Delta\mu$ vanishes periodically.
From Eq.\ (\ref{Illumin_Sig}), one sees that $\Delta\sigma(B)$ is proportional
to $\Delta\mu$, so that they exhibit synchronous oscillations.
For $\vert E_{\mathrm{F}}\vert\gg\hbar\omega_{c}$, if we neglect the oscillations
 in $\Delta\sigma(B)$, by assuming $E_{F}$ not very close to
 $\mbox{sgn}(n)\sqrt{2\vert n\vert}\hbar\omega_{c}$ with $n\neq 0$,
 we can replace the summation over $n$ in $\varTheta$ by an integral $
\sum_{n=1}^{n_{c}}
\rightarrow \int_{0}^{n_{c}} dn$, and then obtain $\varTheta\simeq 2(\hbar\omega_{c}/E_{F})^2$.
In this case, Eq.\ (\ref{JxA}) reduces to
\begin{align}
\Delta\sigma(B)  &  =\frac{e^2}{4\pi^{2}\hbar}\frac{(eB)^2\upsilon_{F}^2}{E_{F}^2}
\upsilon_{F}\tau_{inter}\ ,\label{sigA}
\end{align}
which recovers the classical formula obtained in
Refs.~\cite{PhysRevB.88.104412,PhysRevB.91.245157}, and its result is also plotted
in Fig.\ \ref{Figmagto}(c) as a red dashed line.
The classical formula does not show any oscillations, and
is approximately consistent with the envelope
of the quantum formula for $\vert E_{F}\vert\gg\hbar\omega_{c}$.
For relatively strong magnetic field,
$\vert E_{F}\vert\lesssim\hbar\omega_{c}$, however,
the two formulas deviate
from each other substantially. In the strong-field regime,
for $\vert E_{F}\vert<\sqrt{2}\hbar\omega_{c}$, where
the Fermi level crosses only the $n=0$ LLs, we have $n_{c}=0$
and $\varTheta=1$. In this limiting case, the quantum formula\ (\ref{JxA}) can be simplified to
\begin{align}
\Delta\sigma(B)    =\frac{e^{2}}{2\pi^2\hbar}\frac{eB\upsilon_{F}\tau_{inter}
}{\hbar}\ .\label{JxA_strong}
\end{align}
This formula is in agreement with that derived in Ref.~\cite{PhysRevB.88.104412}
in the ultra-quantum limit (except for an extra prefactor $1/2$ in the latter).
The unsaturated LMC  becomes linearly scaled with $B$, being independent of $E_{F}$.


\begin{figure}[ptb]
\includegraphics[scale=0.7]{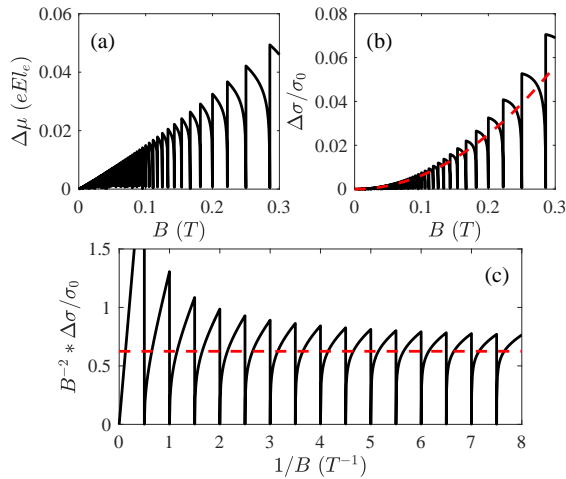}
\caption{ (a) $\Delta\mu$, and (b) $\Delta\sigma(D)$ versus the magnetic field $B$.
 (c) The data of (b) are replotted  to show the periodic-in-$1/B$ dependence of $\Delta\sigma(B)$.
 The red dashed lines in (b) and (c) are obtained from classical formula\ (\ref{sigA}).
 Here, $\sigma_{0}$ is defined in Fig.\ 2, and the Fermi energy is set to $E_{F}=2\hbar\omega_{c}\vert_{B=1T}=2\upsilon_{F}\sqrt{e\hbar[B=1T]}$.}
\label{Figosc}%
\end{figure}
Apart from varying electron Fermi energy $E_{F}$,
the quantum oscillations in the LMC predicted by Eq.\ (\ref{JxA})
can also be  observed in experiments conveniently
by varying the magnetic field $B$.
In Figs.\ \ref{Figosc}(a) and (b), we plot
$\Delta\mu$ and $\Delta\sigma(B)$ as functions of $B$.
We see that both of them oscillate
with $B$. As has been discussed,
both $\Delta\mu$ and $\Delta\sigma(B)$ drop to zero
at $E_{\mathrm{F}
}=\mbox{sgn}(n)\sqrt{2|n|(\hbar\omega_{c})^{2}}$ with $\vert n\vert\neq 0$, or say,
at $1/B=2\vert n\vert e\hbar\left(\upsilon_{F}/E_{F}\right)^{2}$. Therefore,
$\Delta\sigma(B)$ is periodic-in-$1/B$ with the period
\begin{equation}
\Delta\left(\frac{1}{B}\right)=2e\hbar\left(\frac{\upsilon_{\mathrm{F}}}{E_{\mathrm{F}
}}\right)^{2}\ .
\end{equation}
For the parameters chosen in Fig.\ \ref{Figosc}, $\Delta\left({1}/{B}\right)=0.5/T$.
In Fig.\ \ref{Figosc}(c), we replot the LMC
as $B^{-2}\Delta\sigma(B)/\sigma_{0}$ versus $1/B$. We see
that the constant oscillation period is indeed $0.5/T$. Moreover,
the envelope of $\Delta\sigma(B)$ deviates appreciably from the
$B^{2}$ dependence predicted by the classical formula (red dashed line) for
relatively small $1/B$ (large $B$).

In summary, we have derived a unified quantum formula,
Eq.\ (\ref{JxA}), for the chiral-anomaly-induced positive
LMC in WSMs.
It predicts periodic-in-$1/B$ quantum oscillations of the  positive LMC,
as a remarkable fingerprint for identifying a WSM material.
The quantum oscillations of the anomalous LMC are quite
 different from the usual SdH oscillations, the latter appearing in the transverse
conductivity with ${\bf E}\perp{\bf B}$.
The quantum oscillations superposed on the positive LMC are a sufficient condition for identifying a WSM material. In realistic WSMs, if  $E_{F}$ is too far away from the
Weyl points, the LLs essentially become semi-continuous  and impurity scattering may
obscure the quantum oscillations, especially at low fields. However, it is expected that the
quantum oscillations, as an intrinsic property of the anomalous LMC,
will emerge in high-quality WSMs at sufficiently strong
magnetic fields and low temperatures.

\section{Acknowledgments}

This work was supported by the State Key Program for Basic Researches of China
under Grants No. 2015CB921202, No. 2014CB921103 (L.S.), and No. 2017YFA0303203 (D.Y.X),  the National
Natural Science Foundation of China under Grants No. 11674160 (L.S.), No.
11574155 (R.M.), No. 11474149 (R.S.).


\end{document}